\documentstyle[preprint,oldlfont,aps,doublespace]{revtex}
\begin{document}
\draft
\title{Theory of Kinetic Partitioning in Protein Folding
\newline (With Application to  Prions)} 
\author{V. I. Abkevich, A. M. Gutin, and E. I. Shakhnovich}
\address{Harvard University, Department of Chemistry\\
12 Oxford Street, Cambridge MA 02138}
\date{\today}
\maketitle
\begin{center}

\vspace{30pt}

Correspondence to E.Shakhnovich \\
Dept of Chemistry Harvard University\\
12 Oxford Street Cambridge, MA 02138\\
tel 617-495-4130 \\
FAX 617-496-5948 \\
Internet: eugene@diamond.harvard.edu \\
\vspace{30pt}

Keywords: prion protein; conformational transition; protein folding \\
\end{center}
\newpage
\begin{abstract}
In this paper we study the 
phenomenon of kinetic partitioning
when a polypeptide chain has 
two ground state conformations one
of which is more kinetically 
reachable than the other.
This question is relevant to 
understand the phenomenology of prions,
proteins which exist in the 
cell in non-pathogenic $\alpha$-helical
conformation but under certain 
circumstances may transform into
pathogenic $PrP^{Sc}$ state 
featuring increased $\beta$-sheet content.
We designed sequences for lattice model proteins 
having two different conformations
of equal energy corresponding to the 
global energy minimum. 
Folding simulations revealed that one of these conformations
was indeed much more kinetically accessible than the other.
We found that the number and strength 
of local contacts in the ground state conformation is the major factor
which determines which conformation is reached faster:
The greater the number of local contacts the more kinetically reachable
a conformation is. We present simple statistical-mechanical arguments
to explain these findings. The presented results are in a 
clear agreement
with experimental data on prions and 
other proteins exhibiting
kinetic partitioning.
\end{abstract}

\newpage
\section*{Introduction}
The study of prion biology and diseases is a new emerging area of 
biomedical investigation (reviewed by Prusiner, 1992). Prions are proteinaceous 
infectious particles that are composed largely, if 
not entirely, of an abnormal form of the prion protein (PrP) designated, in the
case of scrapie, PrP$^{Sc}$ (Prusiner, 1991). The term 
``prion'' was introduced by Prusiner (1982, 1991), who has shown that prions 
are unique among all infectious pathogenesis 
may be both inherited and transmissible. 
Six diseases of animals and four of humans 
are caused by prions (Prusiner, 1992). The initial hypothesis was that PrP$^{Sc}$
was derived from normal cellular prion protein (PrP$^C$) by a 
post-translational process(Borchelt {\em et al.}, 1990, 1992; Caughey \& Raymond, 
1991; Taraboulos {\em et al.}, 1992). Attempts to identify a post-translational 
chemical modification have been unsuccessful (Stahl {\em et al.}, 1993). Recent 
structural studies demonstrate that PrP$^{Sc}$ and PrP$^C$ have the 
same chemical structure but dramatically 
different conformations (Pan {\em et al.}, 1993). This 
suggests that prions have two low free energy 
states in both of which they can be stable during the lifetime of a 
protein, with pathogenic form PrP$^{Sc}$ being much less soluble
than than cellular PrP$^C$.
The prion puzzle has two aspects. First, it is the reason and the
character of transition from cellular to the pathogenic form.
The most intriguing aspect of it is unusual infectivity 
by PrP$^{Sc}$
which clearly suggests that intermolecular interactions are likely to 
play 
an important role here.  A 
possible mechanism of transition from PrP$^{C}$ to PrP$^{Sc}$ has been suggested recently (Cohen {\em et al.}, 1994) and its implications for prion pathogenesis 
were discussed. 

The focus of this paper, however, is on another important 
aspect of
prion puzzle, namely: Why after 
(or concomitantly with) 
the synthesis of their primary 
structure, prions  do 
not fold into the  
conformation of PrP$^{Sc}$, at the first hand?

One (trivial) explanation may be that 
PrP$^{C}$ is much more thermodynamically stable
than monomeric PrP$^{Sc}$. However, the experimental
data do not seem to support this view. One (indirect) 
evidence is that PrP$^{C}$ is much more 
susceptible to proteolysis 
than PrP$^{Sc}$. Pan {\em et al} (1993)
suggested that PrP$^{C}$ is a kinetic trap.

If PrP$^{C}$ and monomeric
PrP$^{Sc}$ have comparable thermodynamic stabilities
and the partitioning into normal and pathogenic conformations
was driven by thermodynamic rules,
the yield of pathogenic species after 
synthesis of primary structure should be large. 
However, this is not so, and dominant
form in normal, uninfected cells is 
PrP$^C$. Apparently, 
this may be so because PrP$^C$ conformation
is more reachable kinetically. 
 This links prion phenomenology
to the most fundamental aspects of protein folding
since they provide a clear (but not unique, see below)
example of kinetic partitioning when
a conformation becomes dominantly populated
for kinetic, rather than thermodynamic, reasons.

In this paper, we, intrigued by the mystery
of prions, study the phenomenon of kinetic partitioning
in protein folding. Currently, folding 
simulations are feasible only in the realm
of simplified
lattice and off-lattice models,
which already provided useful, experimentally testable,
insights
into such general principles of folding 
as nucleation mechanism (Abkevich {\em et al.}, 1994b; 
Shakhnovich {\em et al.}, 1996;
Fersht, 1995; Itzhaki {\em et al.}, 1995) and folding intermediates 
(Gutin {\em et al.}, 1995a; Mirny {\em et al.}, 1996; Fersht, 1995).

Our approach to study kinetic partitioning, relevant to prion folding,
aims to simulate it
in a simple, yet, nontrivial, model and seek full understanding
of this phenomenon
within the framework of the model.
Then using the model results as an example where kinetic partitioning
is understood, one can apply this as a lead to thinking
about real systems and as a tool to understand existing
and plan future experiments.
Of course such approach may be useful only if there is a generic 
reason for kinetic partitioning, which may be applicable not only to prions,
but to a wider range of systems. In this paper we reveal
a possible physical mechanism of kinetic partitioning which is
due to global structural 
properties of native conformations and we argue that
it may explain certain observed 
features of prion folding as well
as other proteins where kinetic
partitioning is likely to be important. 

In our model study we would like to focus on the 
kinetic aspect of prion folding. 
Therefore, while it is not known which conformation,
PrP$^C$ or PrP$^{Sc}$, has lower free energy,
in the model, we assume that two ``native'' conformations
have equal (free) energies and seek the reason why one of them
is more reachable than the other.

Specifically, in this work we design sequences having two 
different conformations of global energy
minimum and study their folding. 
Our results provide structural clues to 
kinetic partitioning explaining, why in a system with two
ground state conformations, 
one of them 
can be more kinetically reachable than
the other.

We compare the results of our study with phenomenology of prions
and other proteins having kinetic partitioning.

\section*{The Model}
We represent proteins as self-avoiding chains on an infinite cubic
lattice, such that covalently linked residues occupy neighboring
lattice sites.  The energy of a 
conformation is the sum of 
energies of pairwise contacts 
between monomers.  Two monomers are
defined to be in a contact if they are 
neighbors on the lattice and not
connected by a covalent bond.  The energy of a 
contact depends only
on the identity of the two amino acids involved.  The interaction
energies for amino acid pairs are determined from the statistical
distribution of contacts in real proteins (Miyazawa \& Jernigan,
1985, Table VI). 

Our approach to folding simulations requires first to choose a target 
conformation to be the native and design
amino acid sequences that 
fold to, and are stable in, this conformation. It was shown previously
(Goldstein {\em et al.}, 1992; Shakhnovich \& Gutin, 1993; Gutin {\em et al.}, 1995b) that 
such a sequence should render the target native conformation as a 
pronounced global energy minimum. This is the criterion used in our design
algorithm (a Monte Carlo optimization in sequence space).  The details
of this algorithm have been published elsewhere 
(Abkevich {\em et al.}, 1995). 
The quantity that is minimized in this design is 
relative value of native energy comparatively to 
the non-native conformations 
(Bowie {\em et al.}, 1991; Gutin {\em et al.}, 1995b):
\begin{equation}
Z= \frac{E_{nat}-E_{av}}{\sigma}
\end{equation}
where $E_{nat}$ is energy of the native 
conformation, $E_{av}$ is the average 
energy of compact non-native conformations 
with corresponding dispersion 
$\sigma$.

Our aim in this work is to design sequences
that  exhibit prion-like behavior, i.e. which have more than
one conformation as a global energy minimum.
Correspondingly, the design procedure should be modified
to achieve that goal.
To this end we choose two different conformations and 
design such amino acid sequences that have low energy
in both conformations. The essential parameter 
which is minimized in this design is
\begin{equation}
Z = Z_1 + Z_2 + (Z_1 - Z_2)^2
\end{equation}
where $Z_1$ and $Z_2$ are relative energies of the model polypeptide in the 
first and second target conformations, respectfully. The last term in the 
equation
(2) is introduced to ensure that designed sequence have close energies in target conformations.

\section*{Results}
The first model protein we studied 
is a 36-mer chain which has equally low energy in
two target conformations (Fig.1) arbitrarily 
chosen out of more than 84 million possible 
fully compact conformations (Pande {\em et al.}, 1994).
\marginpar{Fig.1}
The sequence was designed to have low 
energy in both of these
conformations (Fig.1c)\footnote{One may notice that sequences shown on Fig.1c and Fig.2c have unusually large content of W and M. This is due to the fact that our design program does
not take into account geometrical properties of amino acids and their natural 
occurrence but only energy of interactions between them, which is determined only approximately. The strongest hydrophobic interactions in the energy set
used in this work (Miyazawa \& Jernigan, 1985, Table VI) is a one between tryptophan and methionine. That is why these amino acids dominate in our designed sequences.}. 

In order to simulate folding, we used the 
standard Monte Carlo method (Hilhorst \& Deutch, 1975). 
Different simulation runs begin from different random coil conformations. 
In the process of 
simulation designed protein always
reached each of the conformations 
shown in Fig.1. Further,
in a long Monte Carlo folding run ($10^9$ steps) 
we did not observe any conformations having
energy lower than these two which 
is indicative that these conformations
are the lowest in energy. 

The mean first passage time (MFPT) \footnote{MFPT in all experiments
reported  in this paper was calculated by 
averaging over 50 folding runs each starting from different random 
conformations.} into the structure shown in Fig.1b is found 
at least 20 times longer than into the one shown Fig.1a in a wide range of 
temperatures. To verify this result we measured MFPT for ten nonhomologous designed 
sequences that have close low 
energies in both conformations shown in Fig.1 
and we consistently 
observed much faster folding into the 
conformation shown in Fig.1a. 
This result was unexpected, and the first guess
was that one of the conformations shown in 
Fig.1 is more kinetically reachable for geometric,
or algorithmic reasons. To test this hypothesis we designed 
sequences were designed having each of the structures shown on 
Fig.1 as their {\em unique} native conformation. MFPT for these sequences 
at their folding transition temperature into their respective 
native conformations were 
approximately equal (data not shown). This ruled out the
simple, but mostly artificial, possibility mentioned above. 

Difference in folding time into conformations with completely different structure
but close energy was also observed 
for random model polypeptides. 
Studying folding of 
ten random 27-mers with the same amino acid composition we found that nine of them fold into their lowest energy conformation in about $10^6$ 
Monte Carlo steps. 
However, the tenth sequence folded 
into the conformation with the lowest energy 
in more then $10^8$ steps. However folding 
time at the same temperature into a completely 
different conformation with only slightly 
higher energy is 20 times faster. 
Again such a difference in  folding times appeared to be
not just due to geometrical inaccessibility 
of the conformation with the lowest 
energy. Folding of the sequence which was 
designed to have this conformation as the global energy minimum was fast.

It seems that when random or specially designed 
sequence have close energy in two conformations folding time into these 
conformations is often quite different. 
This can be similar to what is found for prions which fold 
into their native conformation and 
stay in it during life time of a 
protein being digested by a protease 
before reaching the alternative stable
conformation.  
 
How can the existence of one low energy state influence folding into 
another? The assertion that one conformation can
play a role of a kinetical trap (Abkevich {\em et al.}, 1994a) is not 
sufficient to explain
the observed behavior. At studied temperatures random 27-mer is 
not stable in its low energy conformation, 
 and polypeptide spends most of 
its time in the unfolded state. Since 
the difference in folding times remains 
even when the ground state is unstable 
one should seek  the reason of 
this difference in the propoerties of the 
unfolded state. 
It is known that in unfolded state strong 
local contacts prevail (DeGennes, 1979; 
Grosberg \& Khohlov, 1994). 
The ground state conformations 
are the lowest in energy and 
thus ground state contacts are the 
strongest on average. So among contacts even 
in a unfolded state those which are 
the same as in the ground state, especially 
local ones, will dominate. Comparing 
the structure shown in Fig.1b with the 
one shown in Fig.1a we found significant 
difference in the
number of local contacts (contacts between 
monomers $i$ and $i + 3$). There are 
six local contacts  in the structure 
shown on Fig.1b and eleven such contacts 
in the structure shown on Fig.1a.

When we compared the lowest energy state 
of a random 27-mer, for which kinetic
partitioning was observed, with the alternative low-energy
conformation, which was reached much faster, we
found that both conformations have the same
 number of local contacts. 
 However,  average 
energy of local contacts was lower in the conformation 
faster reachable conformation 
(-0.27) than in the native
one (-0.17). 

Thus one can suggest that the number and strength of local contacts
are the major factors determining which conformation will be reached first. To
pursue this lead further, we chose another couple of structures for 
design: the one with many (seventeen) local contacts (Fig.2a)
\marginpar{Fig.2}
and another one without local contacts (Fig.2b). A sequence 
designed to have equally low 
energy in these structures is shown on Fig.2c.If our hypothesis is 
correct, this sequence   will fold into the structure shown on Fig.2a faster than into the one shown on Fig.2b, 
and the ratio of MFPT should be greater than 
 for structures shown on Fig.1. The results of simulations 
fully confirm this prediction: At temperature when folding is the
fastest, the MFPT from random coil
into the conformation shown on Fig.2b is equal 
to $5.6\cdot10^6$ Monte Carlo steps, 
whereas MFPT into the conformation shown 
on Fig.2b is equal to $7.8\cdot10^8$: more
than 100 times longer. 

To show that this result is generic and 
does not depend on the parameter set, 
we extended analysis using so 
called Go model (Taketomi {\em et al.}, 1975) in which 
all native interactions are set equal and attractive and all other 
interactions are set to zero. Apparently, such model does not consider any 
interactions except the native ones, 
and hence it is somewhat unphysical. However
the Go model has some important advantages 
for the problem under investigation. Firstly,
all contacts  in both ground states have the same energy, and thus 
total energy of ground state 
contacts is simply proportional to their number. 
Secondly, many factors which
influence folding rate in more realistic sequence models 
(dispersion of energy of native contacts (Abkevich {\em et al.}, 1996), 
stability of the specific nucleus (Abkevich {\em et al.}, 1994b; 
Fersht, 1995) and so on)
are not important for the Go model.  Further, the 
MFPT at the optimum temperature for
the Go model is substantially less than for a sequence model. 
This allows one to overcome considerable 
computational difficulties of MFPT determination. 

We compared MFPT for the Go model into the  
conformations shown in Fig.2 at different
temperatures (Fig.3).          
\marginpar{Fig.3}
We found that folding into the 
structure shown in Fig.2a is at least 100 times 
faster than into the one shown on Fig.2b. 

The analysis of a typical folding trajectory for a long 
Monte Carlo run at folding transition temperature (Fig.4) provides 
further insight into the origin of such a pronounced difference
in folding rates into two different ground state structures. 
\marginpar{Fig.4}

For convenience of subsequent discussion we denote
the structure with many local contacts (Fig.1a, Fig.2a)
as $N_1$ and the structure with a few 
local contacts (Fig.1b, Fig.2b) as $N_2$. 
It can be seen from Fig.4 that 
properties of unfolded state 
give rise to the differences 
in folding into $N_1$ and $N_2$.
The chain rapidly transforms into the
set of (unfolded) states with 
significant structural similarity
with $N_1$, having about 50\% of 
the contacts in common with this ground state 
conformation. In contrast, the 
unfolded conformations bear very 
little structural similarity with $N_2$ and 
only after almost $3\cdot 10^7$
MC steps it reaches conformations with sufficient number (and location)
of $N_2$ contacts  to enable nucleation and subsequent
rapid folding into that ground state conformation. This is due to the fact that local contacts, numerous in $N_1$, are more favorable than other possible 
contacts. After these contacts are formed, structural 
similarity with $N_1$ is enforced. This facilitates faster 
folding into $N_1$. In order to nucleate folding into $N_2$, the most 
stable local contacts in the {\em unfolded} state must be broken. This process
is uphill in free energy and therefore requires longer time.  

This qualitative picture can be represented by the following scheme:

\hspace{185pt} $U_1 \Longleftrightarrow N_1$

\hspace{185pt} $\Updownarrow$

\hspace{185pt} $U_2 \Longleftrightarrow N_2$

Here $U_1$ and $U_2$ - unfolded states in which dominate contacts from the corresponding ground states. $U_1$ might be similar to $PrP^*$ state (Cohen {\em et al.}, 1994). 
As was noted earlier, folding from the state $U_1$ into 
the state $N_2$ firstly requires to tear contacts in $U_1$ which are common 
with $N_1$ but not with $N_2$. This corresponds to the transition from $U_1$ 
into $U_2$. The same is also true for the states $U_2$ and $N_1$. 
The essential features of the free energy landscape of prions are 
summarized on a
schematic  diagram  shown on Fig.5.
\marginpar{Fig.5} 

Even if the rate of folding from $U_1$ to $N_1$ is equal to such from 
$U_2$ to $N_2$\footnote{This is a reasonable assumption 
because folding rates of sequences, having only one of the 
conformations shown in Figs.1 and 2 as native 
are close to each other.}, the folding time 
from the random coil state into states $N_1$ and 
$N_2$ will be different if equilibrium constant $K = [U_1]/[U_2]$ between 
states $U_1$ and $U_2$ is not unity. 
If $K \ll 1$ it can be easily shown that MFPT into the state $N_1$ is 
$K \cdot (K_1 + 1)$ times faster then MFPT into the state $N_2$. Here 
$K_1 = [N_1]/[U_1]$ is the equilibrium 
constant between states $N_1$ and $U_1$. When the 
ground state is stable ($K \gg 1$) the ratio of 
folding rates into states $N_1$ and $N_2$ is gretaer 
than in the case when ground state is unstable. This is due to the 
fact that at such conditions the state $N_1$ should be considered as a kinetic trap 
for folding into the state $N_2$.

This conclusion is qualitatively consistent with our numeric results
which suggest (Fig.3) that 
the ratio of MFPT into ground state conformations becomes greater 
as temperature
decreases, i.e. both ground states become thermodynamically stable. To
test this prediction quantitatively, we estimated the 
temperature dependence of 
the equilibrium constant $K_1$ using histogram 
technique (Ferrenberg and Swendsen, 1989; Sali {\em et al.}, 1994; 
Socci and Onuchic, 1995;
Abkevich {\em et al.}, 1995). A long Monte
Carlo simulation was performed ($2\cdot10^7$ steps) 
in which the conformation shown in 
Fig.2a was folded and unfolded many times ($\sim 100$) but the 
conformation shown on Fig.2b was not yet reached. 
The statistics of occurrence of different states were collected. The two important parameters were taken into account: the 
energy of the chain $E$ and similarity parameter $Q_1$. The quantity of 
interest is the logarithm of the density of states $\nu(E,Q_1)$. Once 
calculated, it makes it possible to calculate population of states at different 
temperatures. This provided the estimate for the 
temperature dependence of 
equilibrium constant $K_1$. Fig.6 
\marginpar{Fig.6} 
shows that the ratios of MFPT into states $N_1$ and $N_2$ at different temperatures fit theoretical curve quite well. 

We have shown that number and strength of local contacts
is an important factor which determines which of the two
global minimum conformations is reached faster. This effect was observed for 
the model with designed and random sequences and for the Go model. 
However the question remains: whether strength and number of local 
contacts is the only factor which determines the ratio of MFPT into 
ground state conformations. To address this issue, we randomly selected two 
maximally compact conformations with equal number of local contacts (six).  
To  ensure that energy of local ground state contacts is the same for both 
structures, we studied their folding using the Go model. We found that folding 
into one these structures is somewhat faster than into another.
However, the ratio of MFPT into 
ground state conformations is significantly smaller than 
what we observed 
for the structures shown on Figs.1 and 2 (the data not shown). 
This ratio never exceeds a factor of three and at some 
temperatures is as small as 30\%. 
Such  numbers are typical scatter of MFPT
for model proteins with 
different but unique native states. 

This result suggests that for the studied model the number and strength of 
local contacts in the ground state is the only major factor which 
determines which of the ground state conformations  is reached faster. 
      
\section*{Discussion}
It is important to compare the  properties of natural 
prions with model polypeptides with degenerate 
ground state which we studied in the present work. Structural studies demonstrate that PrP$^C$
and PrP$^{Sc}$ differ in conformation (Pan {\em et al.}, 1993). PrP$^C$ was found to
have high content of $\alpha$-helix (42\%) and essentially no $\beta$-sheet (3\%), whereas PrP$^{Sc}$ had a $\beta$-sheet content 43\% and an 
$\alpha$-helix content 30\%. This is consistent with our findings because in a
$\beta$-sheet non-local contacts dominate, whereas in an $\alpha$-helix local 
contacts dominate.

The fact that PrP$^{Sc}$ is much less soluble than
PrP$^C$ has been well established  
(Pan {\em et al.}, 1993, Kocisko {\em et al.}, 1994). As pointed out in the
Introduction, one can suggest
a ``thermodynamic'' explanation of prion behavior, namely that the stability
of monomeric PrP$^{Sc}$ is vanishingly low. The implication
of this possibility will be that under no circumstances,
including any dilution or mildly denaturing conditions
can monomeric PrP$^{Sc}$ be observed. This seems to contradict
the experimental finding that dilution of prion
solution from 3M of denaturant preserves
infectivity while dilution from 6M of denaturant 
eliminates it
(Kocisko {\em et al.}, 1994). However interpretation 
of these experiments  is not entirely clear (e.g. whether dilution
from 6M of GuHCl leads to folding into PrP$^C$ or not) and we cannot
completely rule out the ``thermodynamic'' explanation though 
we find it less likely.
Perhaps further experiments including the stopped flow folding
of prions upon rapid dilution from 6M GuHCl with subsequent 
monitoring of secondary structure  formation
can be useful in clarifying this very important
question.

Prions are not the only proteins which can spontaneously 
undergo global structural changes
and transform into another stable state. 
The theory described here predicts 
that in such transitions fraction of local 
contacts should decrease. What can be  
observed experimentally is the decrease 
of content of $\alpha$-helixes accompanied by 
increase of content of $\beta$-sheets. 

It is important to mention that 
conformational flips from $\beta$-sheet to $\alpha$-helix 
are also observed for some proteins (Reed \& Kinzel, 1993). However these transitions occur 
when protein environment is altered (change of the solvent or addition of 
an agent which stabilizes $\alpha$-helical conformation). In this case
relative stability of native state can change significantly, and the theory 
presented above is not applicable. We should concentrate on the cases when 
proteins are (metha)stable during their life time in different states 
under the same conditions. 
For example, upon exposure to bright illumination
photosystem $II$ irreversibly transforms into a stable inactive state. 
At the same time content of $\alpha$-helix drops from 67\% to 24\% whereas 
content of $\beta$-sheet increases from 9\% to 41\% (He {\em et al.}, 1991). Another example is 
a human plasminogen
activator inhibitor-1 (PAI-1) which also spontaneously folds into a stable inactive
state without cleavage (Katagiri {\em et al.}, 1992). In agreement with the presented theory, inhibitory 
activity of this protein can be restored 
through denaturation and 
renaturation (Hekman \& Loskutoff, 1985; Katagiri {\em et al.}, 1992). 
It was suggested that transition from active into the
latent state is due to transformation of a surface helix into a 
$\beta$-sheet (Mottonen {\em et al.}, 1992).

Another interesting question is the evolutionary origin of prions. It was
estimated that the probability to randomly 
synthesize a protein sequence with 
degenerate {\em stable} ground states by 
chance is low (Gutin \& Shakhnovich, 1993). 
This may imply that even if some properties of the ``abnormal'',
patogenic,  
state seem now useless and 
even harmful, proteins could have been specially 
designed during evolution to have ``abnormal'' state as well as 
the native one. Finding the possible biological 
role of the ``abnormal'' state 
can be important for
understanding of protein's properties. 
Alternatively we can  assume that prion 
proteins have only weakly optimized sequences, 
so that  they are intrinsically
unstable in their ground state conformations 
at physiological temperature. Then probability of accidental 
synthesis of a 
protein with degenerate native 
state is sufficiently high (Gutin \& Shakhnovich, 1993).  

Finally we would like to point out to limitations of
the present analysis.
In order to simulate the effect of kinetic partitioning,
we had to use a simplified model.
Such simplification comes at a price.
One limitation is that 
we studied folding of model proteins  
much shorter than typical experimental
systems exhibiting kinetic partitioning. 
E.g. prions are approximately 200 aminoacids long, while
in the present study we simulated folding of 27-mers and 36-mers.
While much longer model proteins
can be folded on a lattice (successful folding simulations 
of 
model proteins of 
up to 100 aminoacids long have been reported by several
groups (Kolinski {\em et al}, 1993, Shakhnovich, 1994),  the detailed
analysis presented in this paper required thousands of runs 
to collect
sufficient statistics. This is feasible 
only for relatively short model proteins. 
However, we do not think that this is a crucial limitation 
since comparison of our results with experimental situation suggests 
that gross structural features leading to  kinetic partitioning
may well be reproduced by simulations of shorter chains. 

Another  limitation of the present study is  
that it did not include intermolecular
interactions which are important 
for transition from PrP$^{C}$ to PrP$^{Sc}$.
Here we should emphasize once again 
that our results are aimed to explain why 
prion proteins fold into the conformation of 
PrP$^{C}$ after synthesis of their 
primary structure. Certainly, the 
presented simulations and analysis
do not address the 
mechanism of the conformational change of PrP$^{C}$ into PrP$^{Sc}$.
It is possible to simulate small ensemble of lattice chains,
or study template-mediated folding 
and address the issue of prion infectivity and conformational 
transitions (Cohen {\em et al}, 1994). We are planning
to do so in the near future.

\section*{Acknowledgments}
This work was supported by the Packard Foundation
and NIH (Grant GM 52126). 
We thank Fred Cohen for valuable 
comments and Leonid Mirny for useful discussions. 

\newpage
\section*{REFERENCES}
\noindent
Abkevich, V.I., Gutin, A.M. \& Shakhnovich, E.I. (1994a). Free energy 
landscape for protein folding kinetics. intermediates, traps and multiple 
pathways in theory and lattice model simulations. {\em J. Chem. Phys.}, {\bf 101}, 6052-6062.

\noindent
Abkevich, V.I., Gutin, A.M. \& Shakhnovich, E.I. (1994b). Specific nucleus as 
the transition state for protein folding: Evidence from the lattice model. {\em Biochemistry}, {\bf 33}, 10026-10036.

\noindent
Abkevich, V.I., Gutin, A.M. \& Shakhnovich, E.I. (1995). Impact of local and 
non-local interactions on thermodynamics and kinetics of protein folding. {\em J. Mol. Biol.}, {\bf 252}, 460-471.

\noindent
Abkevich, V.I., Gutin, A.M. \& Shakhnovich, E.I. (1996) Improved design of 
stable and fast-folding model proteins. {\em Folding \& Design}, {\bf 1}, 
221-230.

\noindent
Borchelt, D.R., Scott, M., Taraboulos, A., Stahl, N. \& Prusiner, S.B. (1990). Scrapie and cellular prion proteins differ in 
their kinetics of synthesis and topology in cultured cells. {\em J. Cell Biol.}, {\bf 110}, 743-752.

\noindent
Borchelt, D.R., Taraboulos, A. \& Prusiner, S.B. (1992). Evidence for synthesis of scrapie prion proteins in the endocytic pathway. {\em J. Biol. Chem.}, 
{\bf 267}, 6188-6199.

\noindent
Bowie, J.U., Luthy, R. \& Eisenberg, D. (1991). A method to identify protein 
sequences that fold into a known three-dimensional structure. {\em Science}, {\bf 253}, 164-169.

\noindent
Caughey, B. \& Raymond, G.L. (1991). The scrapie-associated form 
of PrP is made from a cell surface precursor that is both protease- and 
phospholipase-sensitive. {\em J. Biol. Chem.}, {\bf 266}, 18217-18223.

\noindent
Cohen, F.E., Pan, K.-M., Huang, Z., Baldwin, M., Fletterick, P.J. \& Prusiner,
S.B. (1994). Structural clues to prion replication. {\em Science}, {\bf 264}, 530-531.

\noindent
DeGennes, P.G. (1979). {\em Scaling concepts in polymer physics}, Cornell University Press, Ithaca, NY.

Ferrenberg,A.M. \& Swendsen,R.H. (1989). Optimized monte carlo data analysis {\em Phys. Rev. Lett.}, {\bf  63}, 1195--1197.

\noindent
Fersht, A.R. (1995). Optimization of rates of protein folding: The 
nucleation-condensation mechanism and its implications. {\em Proc. Natl. Acad. 
Sci. USA}, {\bf 92}, 10869-10873. 

\noindent
Grosberg, A.Yu. \& Khohlov, A.R. (1994). {\em Statistical physics of macromolecules}, AIP Press, New York, NY.

\noindent
Gutin, A.M. \& Shakhnovich, E.I. (1993). Ground state of random copolymers and the discrete random energy model. {\em J. Chem. Phys.}, {\bf 98}, 8174-8177. 

\noindent
Gutin, A.M., Abkevich, V.I. \& Shakhnovich, E.I. (1995a). Is Burst Hydrophobic 
Collapse Necessary for Protein Folding? {\em Biochemistry}, {\bf 34}, 
3066-3076. 

\noindent
Gutin, A.M., Abkevich, V.I. \& Shakhnovich, E.I. (1995b). Evolution-like selection of fast-folding model proteins. {\em Proc. Natl. Acad. Sci. USA}, {\bf 92}, 1282-1286.

\noindent
Hilhorst, H.J. \& Deutch, J.M. (1975). Analysis of monte-carlo results on 
the kinetics of lattice polymer chains with excluded volume. {\em J. Chem. Phys.}, {\bf 63}, 5153-5161.

\noindent
He, W.-Z., Newell, W.R., Haris, P.I., Chapman, D. \& Barber, J. (1991). Protein secondary structure of the isolated Photosystem II reaction center and conformational changes studied by Fourier transform infrared spectroscopy. {\em 
Biochemistry}, {\bf 30}, 4552-4559.

\noindent
Hekman, C.M. \& Loskutoff, D.J. (1985). Endothelial cells produce a latent inhibitor of plasminogen activators that can be activated by denaturants. {\em J. Biol. Chem.}, {\bf 260}, 11581-11587.

\noindent
Itzhaki, L.S., Oltzen, D.E. \& Fersht, A.R. (1995). The structure of the 
transition state for folding of chymotrypsin inhibitor 2 analyzed by protein 
engineering methods: Evidence for a nucleation-condensation mechanism for 
protein folding. {\em J. Mol. Biol.}, {\bf 254}, 260-288.

\noindent
Katagiri, K., Okada, K., Hattori, H. \& Yano, M. (1988). Bovine endothelial cell plasminogen activator inhibitor. Purification and heat activation. {\em Eur. J. Biochem.}, {\bf 176}, 81-87. 

\noindent 
Kocisko, D.A., Come, J.H., Priola, S.A., Chesebro B., Raymond, G.J., Landsbury, 
P.T. \& Caughey, B. (1994). Cell-free formation of protease-resistant prion 
protein. {\em Nature}, {\bf 370}, 471-474.

\noindent
Kolinski,A.  \& Skolnick,J. (1993) A general method for the prediction of three dimensional structure and folding pathway of globular proteins: Application to designed helical proteins. {\em J.Chem.Phys.}, {\bf 98}, 7420-7433.

\noindent
Mirny, L.A., Abkevich, V.I. \& Shakhnovich, E.I. (1996). Universality and 
Diversity of Protein Folding Scenarios: A Comprehensive Analysis with the Aid 
of Lattice Model. {\em Folding \& Design}, {\bf 1}, 103-116.
 
\noindent
Miyazawa, S. \& Jernigan, R. (1985). Estimation of effective interresidue contact 
energies from protein crystal structures: quasi-chemical approximation. {\em Macromolecules}, {\bf 18}, 534-552, Table VI.

\noindent
Mottonen, J., Strand, A., Symersky, J., Sweet, R.M., Danley, D.E., Geoghegan,
K.F., Gerard, R.D. \& Goldsmith, E.J. (1992). Structural basis of latency in plasminogen activator inhibitor-1.  {\em Nature}, {\bf 355}, 270-273.

\noindent
Pan, K.-M., Baldwin, M.A., Nguyen, J., Gasset, M., Serban, A., Groth, D., 
Mehlhorn, I., Huang, Z., Fletterick, R.J., Cohen, F.E., \& Prusiner, S.B. (1993). Conversion of $\alpha$-helixes into $\beta$-sheets 
features in the formation of the scrapie prion proteins. 
{\em Proc. Natl. Acad. Sci. USA}, {\bf 90}, 10962-10966.

\noindent
Pande, V.S., Joerg, C., Grosberg, A.Yu. \& Tanaka, T. (1994). Enumeration of the Hamiltonian walks on a cubic lattice. {\em J. Phys.}, {\bf A27}, 6231-6236.

\noindent
Prusiner, S.B. (1982). New proteinaceous infectious particles cause scrapie.  {\em Science}, {\bf 216}, 136-144.

\noindent
Prusiner, S.B. (1991). Molecular biology of prion diseases. {\em Science}, {\bf 252}, 1515-1522.

\noindent
Prusiner, S.B. (1992). Chemistry and biology of prions. {\em Biochemistry}, {\bf 31}, 12277-12288.

\noindent
Reed, J. \& Kinzel, V. (1993). Primary structure elements responsible for the 
conformational switch in the envelope glycoprotein gp120 from human 
immunodeficiency virus type 1: LPCR is a motif governing folding. {\em Proc. Natl. Acad. Sci. USA}, {\bf 90}, 6761-6765.

\noindent
Sali, A., Shakhnovich, E.I. \& Karplus, M. (1994). How does a protein fold?
{\em Nature}, {\bf 369}, 248-251.

\noindent
Shakhnovich, E.I. (1994). Proteins with selected sequences fold to their unique 
native conformation. {\em Phys. Rev. Lett.}, {\bf 72}, 3907-3909.

\noindent
Shakhnovich, E.I., Abkevich, V.I. \& Ptitsyn, O.B. (1996). Conservative 
Residues and the Mechanism of Protein Folding. {\em Nature}, {\bf 379}, 96-98.

\noindent
Shakhnovich, E.I. \& Gutin, A.M. (1993). Engineering of stable and fast-folding sequences of model proteins. {\em Proc. Natl. Acad. Sci. USA}, {\bf 90}, 7195-7199.

\noindent
Socci,N. \& Onuchic,J. (1995).  Kinetics and thermodynamic analysis of proteinlike heteropolymer: Monte carlo histogram technique {\em J.Chem.Phys.} {\bf   103}, 4732--4744.

\noindent
Stahl, N., Baldwin, M.A., Teplow, D.B., Hood, L., Gibson, B.W., Burlingame, 
A.L. \& Prusiner, S.B. (1993). Structural studies of the scrapie prion protein using mass spectrometry and amino acid sequencing. {\em Biochemistry}, {\bf 32}, 1991-2002.

\noindent
Taketomi, H., Ueda, Y. \& Go, N. (1975). Studies on protein folding, unfolding and fluctuations by computer simulation. {\em Int. J. Peptide Protein Res.}, {\bf 7}, 445-459.

\noindent
Taraboulos, A., Raeber, A.J., Borchelt, D.R., Serban, D. \& Prusiner, S.B. 
(1992). Synthesis and trafficking of prion proteins in 
cultured cells. {\em Mol. Biol. Cell}, {\bf 3}, 851-863.

\noindent
Thirumalai, D. \& Guo, Z. (1995). Nucleation mechanism for protein folding and theoretical predictions for hydrogen-exchange labeling experiments.
{\em Biopolymers}, {\bf 35}, 137-139.
\newpage
\section*{FIGURE CAPTIONS}
\noindent
{\bf Fig.1} (a), (b) Randomly chosen maximally compact 36-mers on a cubic lattice and (c) a sequence for which these conformations have equally low energy. 

\noindent
{\bf Fig.2} Compact 36-mers on a cubic lattice with 
(a) many local contacts, and (b) without local contacts, and (c) a 
sequence for which these conformations have equally low energy. How these 
structures were found is described in our previous work (Abkevich {\em et al.}, 1995).

\noindent 
{\bf Fig.3} Dependence of MFPT for 36-mers 
shown on Fig.2 on the inverse temperature (Go model). Gray points correspond to the folding into conformation shown on Fig.2a, and black points correspond to the
folding into conformation shown on Fig.2b.
 
\noindent 
{\bf Fig.4} Monte-Carlo folding trajectory at folding transition temperature
($T = 0.6$) for the chain which has two ground state conformations shown in Fig.3 (Go model).
a: the MC-Step dependence of the structural similarity $Q_1$ to the conformation
$N_1$ shown in Fig.2a. This parameter was defined as the number of common contacts in the current and the ground state conformation $N_1$ divided by the
total number of contacts in $N_1$. 
b: The same plot but for the conformation $N_2$ shown
in Fig.2b. 
 
\noindent 
{\bf Fig.5} Schematic representation of energy landscape in a case of 
degenerate ground state. $N_1$ and $N_2$ are the ground state conformations, 
$U_1$ and $U_2$ are unfolded conformations in which dominates contacts from the 
corresponding ground state.

\noindent
{\bf Fig.6} Temperature dependence of ratio of MFPT into conformation shown in
Fig.2b (MFPT$_2$) to MFPT into conformation shown in Fig.2a (MFPT$_1$). 
Experimentally observed data (grey dots) is taken from Fig.3. Theoretical curve
is calculated with a help of histogram technique (Sali {\em et al.}, 1994; Abkevich {\em et al.}, 1995). The only fitting parameter was equilibrium constant $K$ which was taken to be equal 81. 

\end{document}